\documentclass[aps,prl,superscriptaddress,twocolumn,longbibliography]{revtex4-2}
\usepackage{amssymb}
\usepackage{graphicx}
\usepackage{amsmath}
\usepackage{braket}
\usepackage[colorlinks=true,linkcolor=blue,citecolor=blue,urlcolor=blue]{hyperref}

\begin{document}

\title{Variational Approach to Quantum Spin Liquid in a Rydberg Atom Simulator}

\author{Yanting Cheng}
\affiliation{Institute of Theoretical Physics and Department of Physics, University of Science and Technology Beijing, Beijing 100083, China}
\affiliation{Institute for Advanced Study, Tsinghua University, Beijing 100084, China}

\author{Chengshu Li}
\affiliation{Institute for Advanced Study, Tsinghua University, Beijing 100084, China}

\author{Hui Zhai}
\email{hzhai@tsinghua.edu.cn}
\affiliation{Institute for Advanced Study, Tsinghua University, Beijing 100084, China}
\date{\today}

\begin{abstract}

Recently the Rydberg blockade effect has been utilized to realize quantum spin liquid on a kagome lattice. Evidence of quantum spin liquid has been obtained experimentally by directly measuring non-local string order. In this paper, we report a BCS-type variational wave function study of the spin liquid state in this model. This wave function is motivated by mapping the Rydberg blockade model to a lattice gauge theory, where the local gauge conservations replace the role of constraints from the Rydberg blockade. We determine the variational parameter from the experimental measurement of the Rydberg atom population. Then we compare the predictions of this deterministic wave function with the experimental measurements of non-local string order. Combining the measurements on both open and closed strings, we extract the fluctuations only associated with the closed-loop as an indicator of the topological order. The prediction from our wave function agrees reasonably well with the experimental data, with only one fitting parameter determined by measurement of Rydberg atom population. Our variational wave function provides a simple and intuitive picture of the quantum spin liquid in this system that can be generalized to similar spin liquid phases in other lattice geometry. \\

\end{abstract}

\maketitle

\noindent
{\bf 1. Introduction} 

Quantum spin liquid (QSL) is an exotic phase of matter where strong quantum fluctuations due to frustrations destroy conventional spin orders, even for the ground state \cite{Anderson,Read,Lee,Balents,Balents2,Zhou,review4}. The efforts of searching the QSL phase have been lasting for decades in condensed matter physics. Various kinds of evidence for QSL have been found in several different materials, and they are provided by measurements such as spin susceptibility, specific heat, and thermal conductivity, although some are still controversial \cite{Zhou,review4}. Recently, experimental evidence of QSL has also been obtained in the programmable quantum simulators of cold Rydberg atoms \cite{Lukin} and superconducting qubits \cite{Google}. The advantage of these systems is that the expectation value of non-local string operators can be directly measured, providing direct access to the topological order of QSL \cite{Ashvin,Lukin}. 

Realizing the QSL state in a Rydberg atom system utilizes the Rydberg blockade effect and coherent coupling between the ground state and the Rydberg excited state of atoms \cite{Ashvin,Lukin}. Previously, the physics of QSL has been studied in various kinds of models, such as the Heisenberg type model in frustrated lattices \cite{Heisenberg1,Heisenberg2,Heisenberg3,Heisenberg4,Balents}, the quantum dimer model \cite{dimer1,dimer2,dimer3,dimer4,Yao1}, the toric code model \cite{toric}, and the Kitaev honeycomb model \cite{Kitaev}. Although the QSL realized in this Rydberg blockade system bears many similarities with the toric code type topological order, the microscopic Hamiltonian of this system is different from all these previously studied models. The Rydberg blockade model of this experiment has been studied numerically by the density-matrix renormalization group (DMRG) method that reveals a $\mathbb{Z}_2$ QSL ground state existing in certain parameter regime, as well as quantum phase transitions from the QSL to trivial states \cite{Ashvin}, and a related model in different lattice has also been studied using DMRG by Ref. \cite{Subir}.

Aside from numerical simulations, the mean-field method and variational approach have been applied to investigate QSL state in various previously studied models \cite{theory1,theory2,theory3,theory4,theory5}. These studies can provide a more intuitive physical picture of the QSL phase. However, these methods have not yet been developed for the QSL in the Rydberg blockade model, although they are highly desirable. This paper reports our results of a variational wave function study for the QSL discovered in this system. We will discuss the intuition for proposing this wave function by mapping the Rydberg blockade model into a lattice gauge theory (LGT), and we will discuss the properties of this wave function and use this wave function to understand the experimental data reported in Ref. \cite{Lukin}.\\

\begin{figure}[t]
    \centering
    \includegraphics[width=0.45\textwidth]{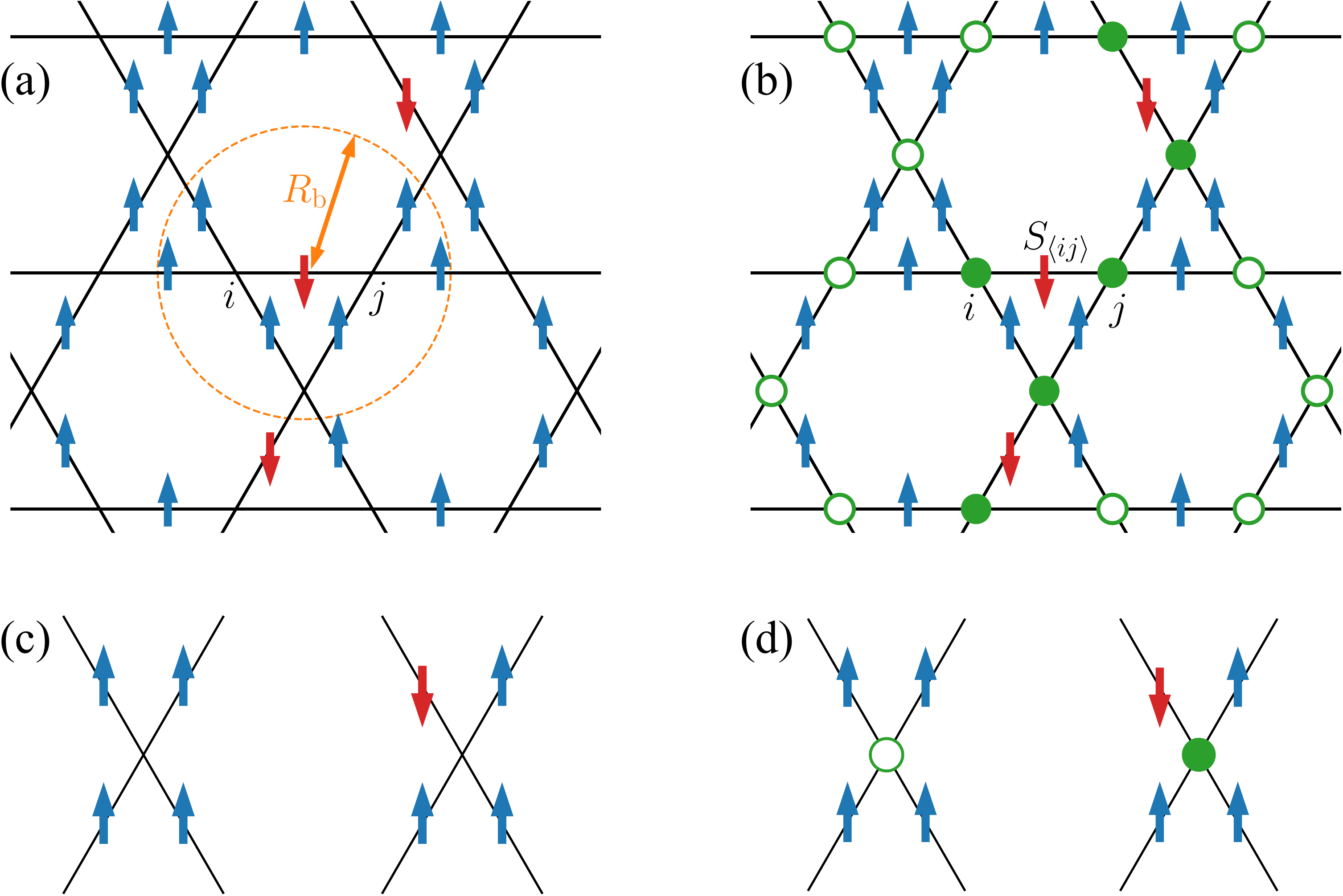}
    \caption{Schematic of the Rydberg blockade model (a) and the equivalent LGT (b). (c) and (d) show the allowed physical configurations at each vertex for the Rydberg blockade model (c) and the equivalent LGT (d). In (b) and (d), open circles denote that the sites are not occupied by $\hat{f}$ fermions and filled circles denote that the sites are occupied by $\hat{f}$ fermions.}
     \label{model}
\end{figure}

\noindent
{\bf 2. Model}

In this system, all atoms are placed at links of a kagome lattice, and a laser field coherently couples atoms between their ground state (denoted by $|\uparrow\rangle$) and the Rydberg excited state (denoted by $|\downarrow\rangle$). The Rydberg blockade radius is shown in Fig.~\ref{model}(a). It means that a Rydberg excited state can block excitation of the other six neighboring atoms inside the circle. This is equivalent to say that for every four atoms sharing one vertex, either all four atoms are in the ground state, or only one out of four atoms is in the Rydberg state, as shown in Fig.~\ref{model}(c). Thus, we can write the Rydberg blockade Hamiltonian as 
\begin{equation}
\hat{H}_\text{RB}=\sum_{\langle ij\rangle}-\Omega\hat{S}^x_{\langle ij\rangle}-\Delta \hat{n}_{\langle ij\rangle}, \label{Rydberg}
\end{equation}         
where $i$ and $j$ are indices of vertex, and summation over $\langle ij\rangle$ denotes summation over all links. $\hat{n}_{\langle ij\rangle}=1/2-\hat{S}^z_{\langle ij\rangle}$ counts the number of Rydberg excitation at the link. $\Omega$ is the coupling strength and $\Delta$ is the detuning. Note that Eq.~\ref{Rydberg} is not a free Hamiltonian because it is subjected to constraints that for each vertex $i$,
\begin{equation}
\sum_{j\in X_i}S^z_{\langle ij\rangle}\geqslant 1, \label{constraint1}
\end{equation}
where $X_i$ denotes four neighboring sites of $i$, and $\sum_{j\in X_i}$ denotes the summation over the four spins sharing the vertex $i$ (see Fig.~\ref{model}(c)). It is clear that negative $\Delta$ favors atoms in the ground state and positive $\Delta$ favors atoms in the Rydberg excited state. For sufficiently large $\Delta$, one of the four atoms sharing each vertex is always in the Rydberg state, and these configurations are called the perfect covering. There are exponentially many perfect covering configurations and the coherent superposition of these configurations gives rise to a QSL state. 

\begin{figure}[t]
    \centering
    \includegraphics[width=0.4\textwidth]{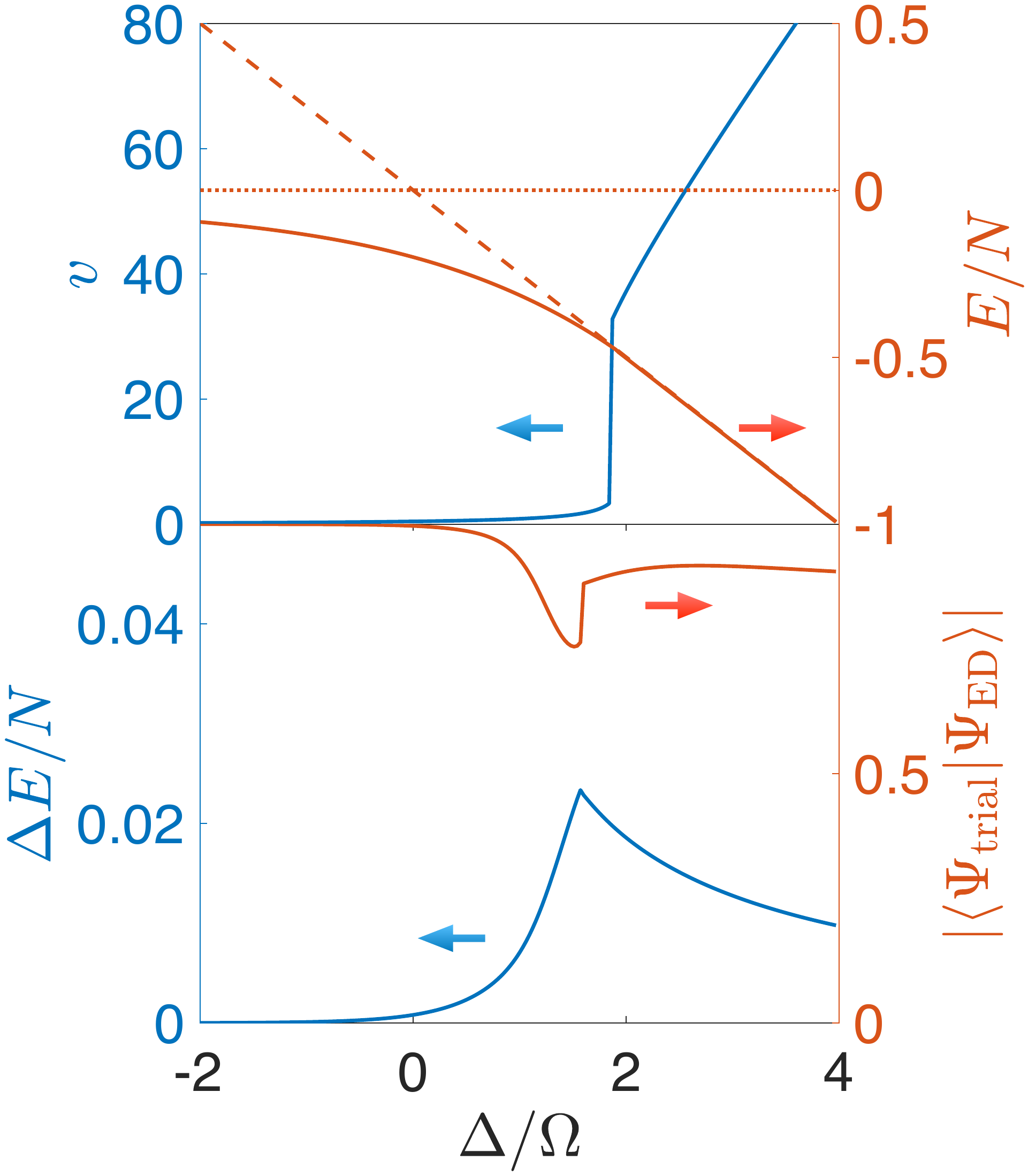}
    \caption{(a) Left axis: The optimum value of $v$ ($u=1$ is fixed) for the variational wave function determined by energy minimization. Right axis: The variational energy $E/N$, compared with the energy of the VBS state (dashed line) and the vacuum state (dotted line). (b) Left axis: The energy difference $\Delta E/N$ between the lowest variational energy and the ground state energy obtained from the exact diagonalization. Right axis: The wave function overlap between the optimum variational wave function and the ground state wave function obtained from exact diagonalization. (a) is calculated for $4\times 4$ unit cells with $96$ sites, and (b) is calculated for $2\times 2$ unit cells with $24$ sites.  }
     \label{variation}
\end{figure}

Now we show that this Rydberg blockade model can be rewritten into an LGT model. To this end, we introduce an auxiliary fermion field $\hat{f}_i$ living on the sites, as shown in Fig.~\ref{model}(b). Then, we write the model as
\begin{equation}
\hat{H}_\text{LGT}=\sum_{\langle ij\rangle}-\Omega\left(\hat{S}^-_{\langle ij\rangle}\hat{f}^\dag_i\hat{f}^\dag_j+\text{h.c.}\right)-\Delta \hat{n}_{\langle ij\rangle} \label{LGT}.
\end{equation} 
In this model, when an atom at link $\langle ij\rangle$ is flipped from $|\uparrow\rangle$ (ground state) to $|\downarrow\rangle$ (Rydberg excited state), two fermions are created at both site-$i$ and $j$. This forbids the spin flip at other links sharing the same vertex-$i$ or $j$. In this way, the Pauli exclusion principle of these auxiliary fermions automatically implements the constraints from the Rydberg blockade. Note that in writing this model, we need to fix an order of fermion pair operators at each link. 

The Hamiltonian Eq.~\ref{LGT} has local gauge symmetries. Let us consider $\hat{f}^\dag_i\rightarrow e^{i\theta_i}\hat{f}^\dag_i$ and simultaneously, $\hat{S}^{-}_{\langle ij\rangle}\rightarrow e^{-i\theta_i}\hat{S}^{-}_{\langle ij\rangle}$ for all four $j\in X_i$, under which Eq.~\ref{LGT} keeps invariant. Thus, these local gauge symmetries ensure local conserved quantities $\hat{Q}_i=\sum_{j\in X_i}\hat{S}^z_{\langle ij\rangle}+\hat{f}^\dag_i\hat{f}_i$. Hence, the inequality constraints Eq.~\ref{constraint1} are translated into equality constraints
\begin{equation}
Q_i=\sum_{j\in X_i}S^z_{\langle ij\rangle}+n_i^f=2,
\end{equation}  
where $n_i^f$ is the number of $f$-fermion at site-$i$. In other word, in the fixed gauge sector $Q_i=2$, this LGT model is equivalent to the Rydberg blockade model. The equivalence between $U(1)$ LGT and the PXP Hamiltonian of one-dimension Rydberg atoms has also been discussed previously \cite{oneD1,oneD2}. This lattice gauge theory is unconventional because of the absence of ordinary electric and magnetic energy terms.  Here the $f$-fermion is reminiscent of the slave particles in usual mean-field theories for QSL, and resembles the role of fractionalization of spin-1 excitation. However, in this case it is introduced in a way different from slave particle techniques.\\

\noindent
{\bf 3. Variational Wave Function}

In the limit of negative enough $\Delta$, all $N$ spins are in the $|\uparrow\rangle$ states and all fermion sites are unoccupied denoted by $\ket{0}$ state, and this state is denoted by $|\text{vac}\rangle=|0\rangle\otimes |\uparrow\rangle^{\otimes N}$. Motivated by this LGT Hamiltonian, we introduce a variational wave function for QSL as
\begin{equation}
|\text{QSL}\rangle=\frac{1}{\mathcal{N}}\prod\limits_{\langle ij\rangle}\left(u+v\hat{f}^\dag_{i}\hat{f}^\dag_j \hat{S}^{-}_{\langle ij\rangle}\right)|\text{vac}\rangle,
\end{equation}
where $\langle ij\rangle$ denotes all links connecting neighboring sites and $\mathcal{N}$ is a normalization factor. In this wave function, the spin flip operator is always combined together with the fermion pair creation operator at each link. Thus, this wave function can be simplified into a BCS type one by only writing its fermion part,   
\begin{equation}
|\text{QSL}\rangle=\frac{1}{\mathcal{N}}\prod\limits_{\langle ij\rangle}\left(u+v\hat{f}^\dag_{i}\hat{f}^\dag_j \right)|0\rangle.  \label{variationalwf}
\end{equation}
Instead, if we return to the spin language, this wave function can be written as
\begin{equation}
|\text{QSL}\rangle\propto\sum_{\Lambda}u^{N^\Lambda_{\uparrow}}v^{N^\Lambda_\downarrow}|\Lambda\rangle, 
\label{variationalwf2}
\end{equation}
where $|\Lambda\rangle$ denotes all allowed physical spin configurations written in the $S_z$ bases, and $N^\Lambda_\sigma$ ($\sigma=\uparrow,\downarrow$) respectively denotes the number of $\uparrow$ or $\downarrow$ spins in the configuration $\Lambda$, with $N^\Lambda_\uparrow+N^\Lambda_\downarrow=N$ fixed.

There are several reasons to consider this variational wave function. First, the BCS type wave function is a very efficient way to represent the superposition of different configurations. Secondly, in the limit $v\rightarrow 0$, the wave function recovers the trivial state with no Rydberg excitations. In the limit of large $v$, as one can see from Eq.~\ref{variationalwf2}, the superposition is dominated by perfect covering configurations with $N^\Lambda_\downarrow=N/4$ where the concentration of monomers approaches zero. Nevertheless, we note that small concentration of monomers assists quantum superposition and plays an important role in establishing the quantum spin liquid. Thus, by varying $v$, the wave function smoothly interpolates the physics in two regimes with sufficiently negative and positive $\Delta$. Thirdly, the spin configurations generated by this wave function automatically satisfy the constraints from the Rydberg blockade. Finally, it is known that the BCS wave function possesses the $\mathbb{Z}_2$ topological order \cite{BCS_Z21,BCS_Z22,BCS_Z23}, and it is natural to use this wave function to describe a $\mathbb{Z}_2$ QSL.  

In practice we can fix $u=1$ and determine $v$ by minimizing the variational energy $\langle \text{QSL}|\hat{H}_\text{LGT}|\text{QSL}\rangle$. The results are shown in Fig.~\ref{variation}. We find that $v$ monotonically increases as $\Delta/\Omega$ increases, and experiences a fast increasing around $\Delta/\Omega\approx1.9$. This feature has been observed in variational calculation on various system sizes. After the fast increasing, $v\gg 1$ and the wave function is nearly an equal weight superposition of perfect covering configurations, recovering the QSL of the toric code type. In this regime, we also observe that the variational energy approaches the energy of the valence bond solid (VBS) state. This is because our model, unlike the quantum dimer model, lacks direct coupling between different perfect covering configurations. The coupling between different configurations is due to higher order processes and is at least of the order of $\Omega^6/\Delta^5$, which is strongly suppressed once $\Delta/\Omega>1$. In the negative $\Delta$ regime, the energy of our wave function approaches the energy of trivial state with no Rydberg excitation from below, as we expected. We have also compared our variational wave function with the exact diagonalization on small system size. Fig.~\ref{variation}(b) shows that the energy difference is at most about $2\%$ of the typical energy scale $\Omega$, and the wave function overlap is typically larger than $0.8$. The largest discrepancy occurs in the regime when $v$ varies very fast as $\Delta/\Omega$ changes.  \\

\begin{figure}[t]
    \centering
    \includegraphics[width=0.48\textwidth]{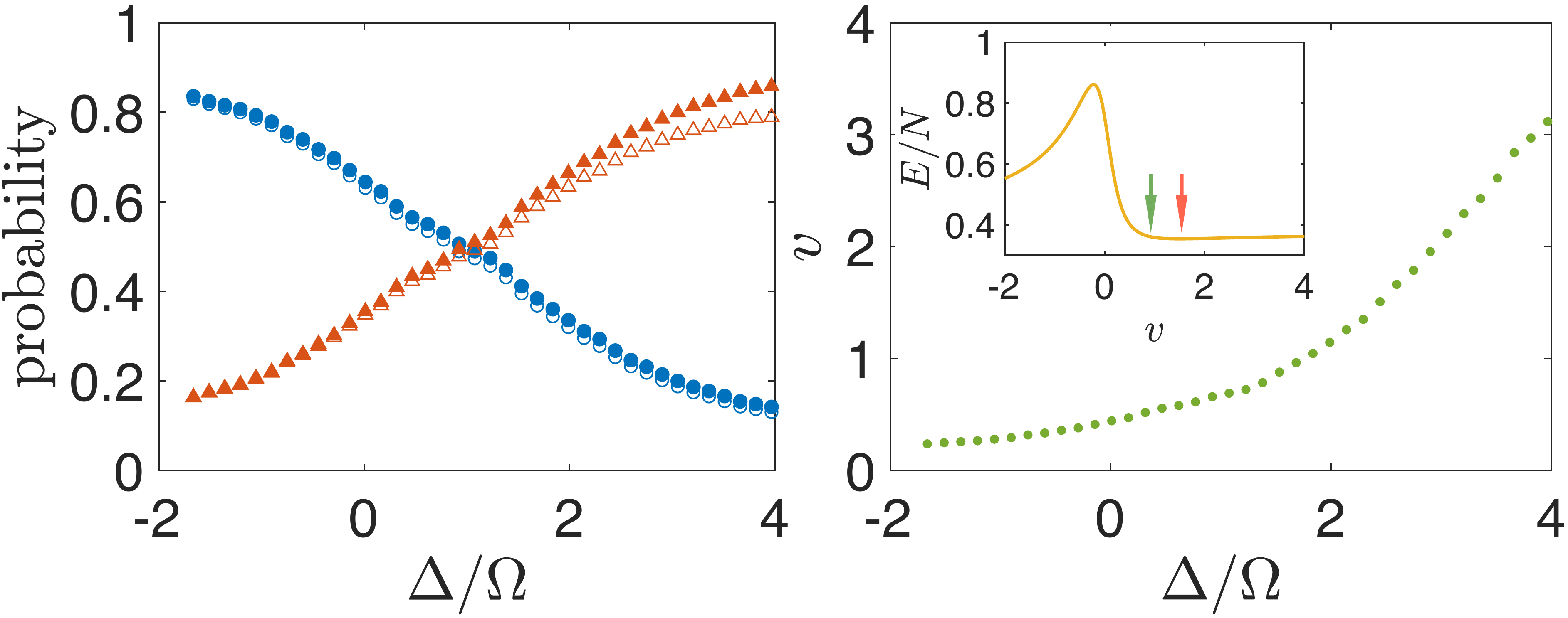}
    \caption{(a) The experimental observation of the percentage of the vertex with no Rydberg excitations (blue circles) and the percentage of the vertex with one Rydberg excitation (red triangles). The open circles and triangles are original experimental data, and the filled circles and triangles are normalized data eliminating the effects of the vertex with two Rydberg excitations. (b) The value of $v$ ($u=1$ fixed) obtained by fitting experimental data. The inset in (b) shows the energy curve as a function of $v$ for $\Delta/\Omega=1.5$. Two arrows mark the fitted and energetically optimum values, whose corresponding energies are very close.}
     \label{exp1}
\end{figure}

\noindent
{\bf 4. Experimental Observations}

Now we use this wave function to explain the experimental data reported in Ref. \cite{Lukin}. Fig.~\ref{model}(c) has shown that there are two cases for each vertex, respectively corresponding to zero or one Rydberg atom out of four atoms linked to each vertex. The percentages of these two cases have been measured experimentally. However, there is a small percentage of cases with two Rydberg atoms sharing the same vertex in practice. Here we ignore this case because they are not captured by our model, and we normalize the experimental data to unity, as shown in Fig.~\ref{exp1}(a).  

\begin{figure}[t]
    \centering
    \includegraphics[width=0.48\textwidth]{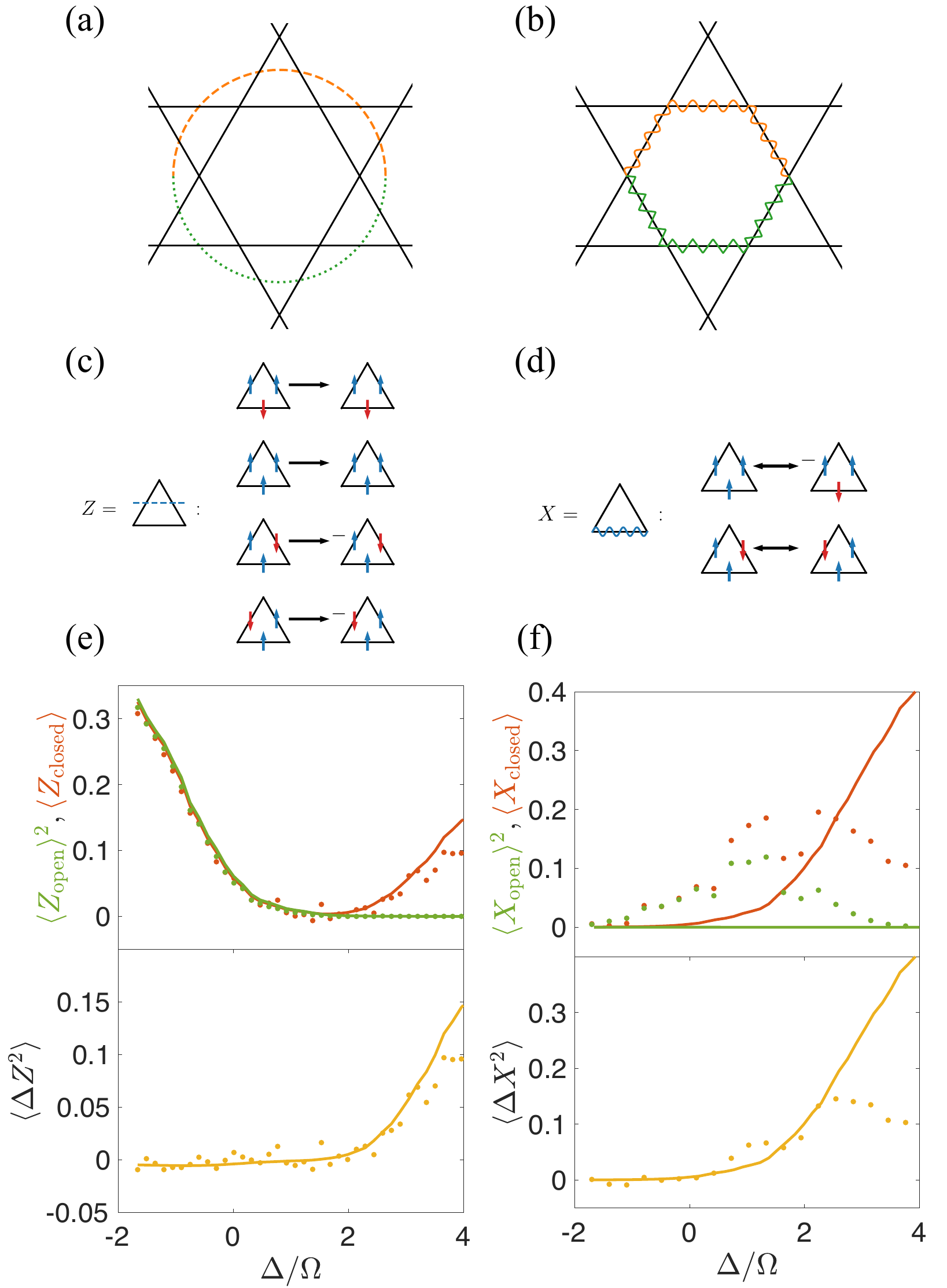}
    \caption{(a, b) Illustration of the closed loop and two open loops (dashed/wavy lines denoting loop-1 and dotted/zig-zag lines denoting loop-2) where the operators $\hat{Z}$ and $\hat{X}$ are introduced in Ref. \cite{Lukin,Ashvin} and their expectation values are measured experimentally in Ref. \cite{Lukin}. (c, d) Illustration of the operators $\hat{Z}$ (c) and $\hat{X}$ (d). (e, f) Comparing the experimental measurements of $\langle\hat{Z}_\text{open}\rangle^2$, $\langle\hat{Z}_\text{closed}\rangle$, $\langle\hat{X}_\text{open}\rangle^2$, $\langle\hat{X}_\text{closed}\rangle$, $\langle\Delta Z^2\rangle=\langle\hat{Z}_\text{closed}\rangle-\langle\hat{Z}_\text{open}\rangle^2$ and $\langle\Delta X^2\rangle=\langle\hat{X}_\text{closed}\rangle-\langle\hat{X}_\text{open}\rangle^2$ with the predictions of the variational wave function, with $v$ given by Fig.~\ref{exp1}(b) ($u=1$ fixed). The dots are experimental data reprinted from Ref. \cite{Lukin} and the solid lines are results from our wave function with parameter $v$ determined by the experimental data of the Rydberg atom population.}
     \label{exp2}
\end{figure}

We require the variational wave function Eq.~\ref{variationalwf} to reproduce these normalized experimental data in Fig.~\ref{exp1}(a). This requirement leads to a value of $v$ for each $\Delta/\Omega$ as shown in Fig.~\ref{exp1}(b). We note that for a given $\Delta/\Omega$, this value of $v$ is significantly smaller than the value obtained from energy minimization. This is because the energy landscape is quite flat in the large $v$ regime, as shown in the inset of Fig.~\ref{exp1}(b). In the experiment of Ref. \cite{Lukin}, $\Delta$ is ramped from negative to a positive value, and therefore, $v$ evolves from a small value to a large value. Since the energy landscape is very flat, it should take a very long time for the system to relax toward the lowest energy state. Thus, it is conceivable that by the time of performing the measurement, the system has not yet reached the lowest energy but has reached the place where the energy landscape is very flat, and the energy of the measured state is already very close to the lowest energy.   

The experimental evidences of QSL are provided by measuring the non-local loop order \cite{Lukin,Ashvin}. Two kinds of loops are shown in Fig.~\ref{exp2}(a, b), and a closed loop is always made of two symmetric open loops, along which operators $\hat{Z}$ and $\hat{X}$ are measured. Here the definitions of $\hat{Z}$ and $\hat{X}$ are introduced in Ref. \cite{Lukin,Ashvin} and are shown in Fig.~\ref{exp2}(c, d), respectively. As shown in Fig. 4(c), $\hat{Z}$ operator is diagonal in the $S_z$ basis. The loop introduced for $\hat{Z}$ crosses links. If there is a Rydberg atom sitting at the link, $\hat{Z}$ operator introduces a minus sign for this wave function. $\hat{X}$ is off-diagonal in the $S_z$ basis and connects different Rydberg configurations as shown in Fig. 4(d). For instance, $\hat{Z}_\text{closed}$ performing operation $\hat{Z}$ along the closed loop shown in Fig.~\ref{exp2}(a).   
Thus, for both operators $\hat{Z}$ and $\hat{X}$, $\hat{Z}_{\text{closed}}=\hat{Z}_{\text{open1}}\hat{Z}_{\text{open2}}$ and $\hat{X}_{\text{closed}}=\hat{X}_{\text{open1}}\hat{X}_{\text{open2}}$. Furthermore, by symmetry, we have $\langle \hat{Z}_{\text{open1}}\rangle=\langle\hat{Z}_{\text{open2}}\rangle\equiv \langle\hat{Z}_{\text{open}}\rangle$, and $\langle \hat{X}_{\text{open1}}\rangle=\langle\hat{X}_{\text{open2}}\rangle\equiv \langle\hat{X}_{\text{open}}\rangle$. 

Hence, we concern whether the order along a closed loop can be fractionalized into products of two orders along two open loops, i.e. whether $\langle\hat{Z}_{\text{closed}}\rangle\approx\langle\hat{Z}_{\text{open1}}\rangle\langle\hat{Z}_{\text{open2}}\rangle=\langle \hat{Z}_{\text{open}}\rangle^2$ and whether $\langle\hat{X}_{\text{closed}}\rangle\approx\langle\hat{X}_{\text{open1}}\rangle\langle\hat{X}_{\text{open2}}\rangle=\langle \hat{X}_{\text{open}}\rangle^2$. In other word, we define 
\begin{align}
&\langle \Delta Z^2\rangle\equiv\langle\hat{Z}_{\text{closed}}\rangle-\langle \hat{Z}_{\text{open}}\rangle^2,\nonumber\\
&\langle \Delta X^2\rangle \equiv\langle\hat{X}_{\text{closed}}\rangle-\langle \hat{X}_{\text{open}}\rangle^2.
\end{align} 
These definitions eliminate the short-range local fluctuations attributed to both open and closed loops, and only signal the the non-local loop fluctuations, revealing the nature of QSL. This bears the same spirits as Fredenhagen and Marcu's normalization \cite{FM1,FM2}. 

In Fig.~\ref{exp2}(e) and (f) we compare the experimental measurements with the predictions from our variational wave function, with $v$ shown in Fig.~\ref{exp1}(b). Note that although $v$ is also obtained from experimental data shown in Fig.~\ref{exp1}(a), these experimental data do not contain any information of quantum coherence or topological orders, and therefore, they are intrinsically different from the data presented in Fig.~\ref{exp2}(e) and (f). When $v$ is fixed by the Rydberg atom population, there is no more variational parameter when computing expectation values of loop operators. It shows that our wave function can well reproduce $\langle \hat{Z}_{\text{closed}}\rangle$, $\langle \hat{Z}_{\text{open}}\rangle$ and $\langle \Delta Z^2\rangle$. However, our wave function cannot reproduce $\langle \hat{X}_{\text{closed}}\rangle$ and  $\langle \Delta X^2\rangle$. Since our wave function is designed to capture long-range coherence of spin configurations, it is conceivable that certain short-range fluctuations are not captured by our wave function. Nevertheless, our wave function captures $\langle \Delta X^2\rangle$ reasonably well up to $\Delta/\Omega\approx 2.2$. For larger $\Delta/\Omega$, the experimental data of $\langle \Delta X^2\rangle$ is smaller than the prediction of our wave function, whose reason requires further investigation. One possible explanation is that the coupling between different configurations is suppressed when $\Delta/\Omega>1$ as discussed above, and the non-local coherence becomes more fragile and more sensitive to noises in the measurements. 

Finally, we note that our variational wave function agrees with measurements of $\hat{Z}$ better than the agreement with measurements of $\hat{X}$. This is probably because our wave function is written on the $S_z$ bases. In this bases, $\hat{Z}$ operator can be easily expressed in terms of our wave function as 
\begin{equation}
\hat{Z}|\text{QSL}\rangle=\frac{1}{\mathcal{N}}\prod\limits_{\langle ij\rangle \in \text{loop}}\left(u-v\hat{f}^\dag_{i}\hat{f}^\dag_j \right) \prod\limits_{\langle ij\rangle^\prime}\left(u+v\hat{f}^\dag_{i}\hat{f}^\dag_j \right)|0\rangle, \label{variationalwfZ}
\end{equation}  
where $\langle ij\rangle \in \text{loop}$ denotes the links crossed by the loop, and $\langle ij\rangle^\prime$ denotes other links. That is to say, the pairing phase changes by $\pi$ at the link crossed by the loop. If the loop crosses a hexagon, pairing phases at two links are changed by $\pi$ and there is no net flux in the hexagon. However, if an open loop ends the center of a hexagon, it creates a $\pi$ vortex in the BCS wave function along this hexagon. This is consistent with our physical understanding of the $m$-anyon in a $\mathbb{Z}_2$ QSL.\\

\noindent
{\bf 5. Summary and Discussion}

In summary, in the Rydberg blockade experiment on a kagome lattice, strong evidence of a QSL has been obtained by measuring the non-local quantum order. We have proposed a variational wave function to understand the QSL state and the consistency of experimental data has been obtained. Although quantitatively less rigorous than the DRMG and the exact diagonalization study, our wave function provides a simple and intuitive picture for understanding this system. This picture can be extended to a similar spin liquid state in other Rydberg array systems by  varying Rydberg blockade conditions and changing lattice geometries. This opens up a new avenue for studying QSL physics.\\  

\noindent
\textit{Note added} There appeared in arXiv several related works \cite{new1,new2,new3} studying the quantum spin liquid in Rydberg atom systems after our paper was submitted.\\

\noindent
{\bf Acknowledgment}

We thank Hong Yao, Yi Zhou, Jingyuan Chen, Chao-Ming Jian, Shang Liu, Zi-Xiang Li and Fan Yang for helpful discussions, and we thank the experimental group of Ref.~\cite{Lukin} for sharing the experimental data. The project is supported by Beijing Outstanding Young Scholar Program, NSFC Grant No.~11734010 and the XPLORER Prize. C. L. is also supported by the International Postdoctoral Exchange Fellowship Program and the Shuimu Tsinghua Scholar Program. Y. C. is also supported by NSFC under Grant No. 12204034.\\

\noindent
{\bf Data availability} The codes for numerical calculations are available at \url{https://github.com/chengshul/QSL_kagome}.

\bibliography{main}

\end{document}